%% file: lipics-v2021-sample-article.tex
\documentclass[a4paper,UKenglish,cleveref, autoref, thm-restate]{lipics-v2021}

\input{preamble_LIPICS}

\title{From Perception to Action: Can UI Interventions Foster Sustainable LLM Chatbot} 

\author{Nitish {Patkar}}{University of Applied Sciences and Arts Northwestern Switzerland (FHNW), Windisch, Switzerland}{nitish.patkar@fhnw.ch}{}{}

\author{Pooja {Rani}}{University of Mannheim, Mannheim, Germany}{pooja.rani@uni-mannheim.de}{}{}

\author{Jack {Gl{\"a}ser}}{University of Applied Sciences and Arts Northwestern Switzerland (FHNW), Windisch, Switzerland}{jack.glaeser@swissfex.ch}{}{}

\author{Simon {L\"uscher}}{University of Applied Sciences and Arts Northwestern Switzerland (FHNW), Windisch, Switzerland}{simon.luescher@easydoo.ch}{}{}

\author{Martin {Kropp}}{University of Applied Sciences and Arts Northwestern Switzerland (FHNW), Windisch, Switzerland}{martin.kropp@fhnw.ch}{}{}

\authorrunning{N. Patkar et al.}

\Copyright{Nitish Patkar and Pooja Rani and Jack Gl{\"a}ser and Simon L{\"u}scher and Martin Kropp}



\EventEditors{John Q. Open and Joan R. Access}
\EventNoEds{2}
\EventLongTitle{42nd Conference on Very Important Topics (CVIT 2016)}
\EventShortTitle{CVIT 2016}
\EventAcronym{CVIT}
\EventYear{2016}
\EventDate{December 24--27, 2016}
\EventLocation{Little Whinging, United Kingdom}
\EventLogo{}
\SeriesVolume{42}
\ArticleNo{23}

\begin{document}

\maketitle

\begin{abstract}
\textbf{Background:}
Large language model (LLM)-powered chatbots are increasingly embedded in everyday workflows, raising sustainability concerns due to their energy use.
Most mitigation strategies emphasize model or infrastructure efficiency, while the user-interface (UI) layer remains underexplored despite its potential to shape interaction behavior.

\textbf{Aims:}
We investigate whether sustainability-oriented UI interventions can increase users' energy awareness and encourage more energy-responsible chatbot use without reducing usability.

\textbf{Method:}
We first conducted a baseline survey with 77 participants to assess awareness and receptiveness to intervention concepts.
Guided by prior work on persuasive technology and choice architecture, we implemented a web-based chatbot prototype with a three-mode switch (Energy-efficient, Balanced, Performance), per-response energy feedback, pre-send energy estimates, a usage metrics dashboard, and energy analogies.
We then evaluated the prototype in a five-day field study with 11 participants.

\textbf{Results:}
In the baseline survey, 94.8\% of respondents reported at least some awareness of AI energy use, yet 88.3\% misestimated actual consumption.
Although concern about environmental impact was high, only 39.0\% indicated willingness to accept a performance trade-off for lower energy use.
In the field study, Energy-efficient mode accounted for 55.8\% of logged prompts (155/278), while 90.9\% (10/11) self-reported actively choosing Eco-mode when high accuracy was not required.
Observed total estimated energy use was 164.72~Wh; under an all-Performance counterfactual, this would increase to 309.44~Wh, implying a 144.72~Wh (46.77\%) lower footprint under observed behavior.
Participants did not reduce prompt length, suggesting mode switching, rather than prompt shortening, as the primary behavioral mechanism.

\textbf{Conclusions:}
Sustainability-oriented UI interventions can improve awareness and support more energy-responsible interaction patterns in LLM chatbots.
These effects are best interpreted as behavioral and model-based estimates that complement backend efficiency work, and the provided prototype and replication package support further research on energy-aware conversational AI design.
\end{abstract}

\input{content}



\bibliography{references}

\appendix

\end{document}

%% file: preamble_LIPICS.tex
\nolinenumbers
\usepackage{amsfonts}
\usepackage{booktabs}
\usepackage{textcomp}
\usepackage{xcolor}
\usepackage{xspace}
\usepackage{ulem}
\usepackage{tablefootnote}
\usepackage{ifthen}
\usepackage[shortlabels]{enumitem}
\usepackage{csvsimple}
\usepackage{supertabular}
\usepackage{tabularx}
\usepackage[most]{tcolorbox}
\usepackage{adjustbox}
\usepackage{pifont}
\usepackage{microtype}
\usepackage{listings}

\newcolumntype{R}[2]{
  >{\adjustbox{angle=#1,lap=\width-(#2),margin*=0.4em 0em 0em 0em}\bgroup}
  l
  <{\egroup}
}

\renewcommand{\arraystretch}{1.2}

\let\oldFootnote\footnote
\newcommand\nextToken\relax
\renewcommand\footnote[1]{%
  \oldFootnote{#1}\futurelet\nextToken\isFootnote}
\newcommand\isFootnote{%
  \ifx\footnote\nextToken\textsuperscript{,}\fi}

\definecolor{gray50}{gray}{.5}
\definecolor{gray40}{gray}{.6}
\definecolor{gray30}{gray}{.7}
\definecolor{gray20}{gray}{.8}
\definecolor{gray10}{gray}{.9}
\definecolor{gray05}{gray}{.95}

\newlength\Linewidth
\def\findlength{
\setlength\Linewidth\linewidth
\addtolength\Linewidth{-4\fboxrule}
\addtolength\Linewidth{-3\fboxsep}
}

\newenvironment{rqbox}{
\par\begingroup
\setlength{\fboxsep}{5pt}
\findlength
\setbox0=\vbox\bgroup
\noindent
\hsize=0.95\linewidth
\begin{minipage}{0.95\linewidth}\normalsize
}{
\end{minipage}
\egroup
\textcolor{gray20}{\fboxsep1.5pt\fbox{
\fboxsep5pt\colorbox{gray05}{\normalcolor\box0}}}
\endgroup
\par\noindent
\normalcolor
\ignorespacesafterend
}

\newcommand{\rb}[1]{
\begin{tcolorbox}[
colback=gray!05,
colframe=black,
width=\linewidth,
arc=3mm,
auto outer arc,
boxrule=0.5pt
]
#1
\end{tcolorbox}
}

\newcounter{finding}

\newcommand{\finding}[1]{%
\refstepcounter{finding}%
\rb{
\noindent
\textit{\textbf{Finding \thefinding.} #1}
}
}

\newcommand\tk[1]{}
\newcommand\ab[1]{}
\newcommand\np[1]{}
\newcommand\pr[1]{}
\newcommand\jz[1]{}
\newcommand\vk[1]{}
\newcommand\lc[1]{}
\newcommand\rC[1]{}
\newcommand\ANS[1]{}
\newcommand\todo[1]{}

\newcommand{\ie}{\textit{i.e.},\xspace}
\newcommand{\eg}{\textit{e.g.},\xspace}

\newcommand{\threeModeSwitch}{\textit{Three-Mode Switch}\xspace}
\newcommand{\metricsDashboard}{\textit{Metrics Dashboard}\xspace}
\newcommand{\promptPrediction}{\textit{Prompt Prediction}\xspace}
\newcommand{\energyNote}{\textit{Energy Note}\xspace}
\newcommand{\energyAnalogies}{\textit{Energy Analogies}\xspace}

\newcommand{\rqI}{What is the current level of user awareness about the energy consumption of LLM chatbots, and how do they perceive sustainability-oriented UI interventions?}
\newcommand{\rqIshort}{User Awareness and Perception}

\newcommand{\rqII}{How do sustainability-oriented UI interventions influence user behavior and interaction patterns?}
\newcommand{\rqIIshort}{Assessing the Impact of UI Interventions}


%% file: content.tex
\section{Introduction}
\label{sec:introduction}
While chatbots have been widely used in customer service for decades now, the emergence of Large Language Model (LLM)-powered chatbots (henceforth LLM chatbots), \eg ChatGPT, Claude, Gemini, and DeepSeek, marked a major shift in their capabilities and societal impact. 
Since the launch of ChatGPT, over 400 million users have been using it weekly~\cite{reuters-openai-400m-users-2025}. 
Organizations from education~\cite{Laba23} to healthcare~\cite{Wah25} are now integrating them or developing custom versions to incorporate domain expertise. 
On the one hand, their rapid growth has enabled a wide range of use cases. 
On the other hand, it has also attracted increasing scrutiny regarding their energy consumption and carbon footprint~\cite{Jian24}, both during training and deployment.

Training state-of-the-art language models demands enormous computational power, which in turn requires substantial electricity and produces significant CO$_2$ emissions. 
For example, training a single large NLP model can emit hundreds of tons of CO$_2$ - an amount comparable to the lifetime emissions of several cars~\cite{Stru19}. 
In response, researchers are calling for ``Green AI'' practices that prioritize energy efficiency and environmental responsibility alongside accuracy or performance~\cite{Verd23,Wu22,Stoj24a,Coig24-preprint,Li24a,Cruz25}.

LLM chatbots consume energy across the full stack and not just during model training, spanning from the back-end model servers to the front-end user interface (UI). 
Yet, most optimization efforts have focused on training and backend optimizations (\eg efficient model architectures, hardware accelerators, or data center improvements)~\cite{Jian24, Li24a, Wilk24a, Stoj25a, Nguy24a,Isaz24}, largely overlooking how the \textit{UI layer}, specifically the user interactions or UI interventions, can influence cumulative energy consumption during everyday use.


Our work shifts this focus to the UI layer. 
\emph{UI interventions} are intentional interface changes that influence users' perceptions or actions~\cite{Sohn15,Hals20}, drawing on persuasive technology and choice architecture~\cite{Fogg09,Muns16,Wein16}.
Prior studies in sustainability contexts show that eco-mode configurations, per-action feedback, and relatable energy analogies can shift behavior while preserving autonomy~\cite{Gron23,Fili23}.
However, these mechanisms have not been studied in LLM chatbot environments, where energy costs are largely invisible and interactions are frequent.
We therefore investigate whether such interventions can increase energy awareness and steer use toward more computationally efficient interaction.
Drawing on prior work on behavior-change mechanisms~\cite{Wein16,Gron23,Fili23}, we identified five candidate UI interventions for study: a \threeModeSwitch{}, a per-response \energyNote{}, a \promptPrediction{}, a \metricsDashboard{}, and \energyAnalogies{} (described in detail in~\autoref{sub:intervention-selection}).
We then followed a three-stage methodology to evaluate them.
First, we conducted an online survey with 77 participants to establish a baseline of user awareness and to confirm receptiveness to these intervention concepts in LLM chatbot contexts.
Second, guided by the survey findings, we implemented the five interventions in a working LLM chatbot prototype.
Third, we evaluated this prototype in a five-day field study with eleven participants.
Our results showed that sustainability-oriented UI interventions increased awareness and encouraged more energy-responsible behavior: 90.9\% reported actively choosing Eco-mode when high accuracy was not required, while usability remained high.
At the same time, \emph{Performance mode} accounted for approximately 89\% of the total energy footprint despite being used in fewer than 25\% of prompts.
These findings highlight that UI design can shift routine behavior, but should complement backend and model-level optimizations for high-cost interactions.

With this work, we make the following contributions:
\begin{itemize}
    \item a survey on the awareness and perception of end users for the energy footprint of LLM chatbots;
    \item empirical evidence on how selected UI interventions influence energy-aware interaction with LLM chatbots; and
    \item a replication package (RP)~\cite{reppackage} containing detailed responses and implemented working LLM chatbot. 
\end{itemize}

\section{Study Design}
\label{sec:study-design}

In this section, we first describe how we identified and selected five candidate UI interventions from prior work (\autoref{sub:intervention-selection}).
We then present the three stages of our methodology: a baseline survey to confirm user receptiveness to the underlying intervention concepts (\autoref{sub:rq1-method}), a prototype implementation of the five interventions in a working chatbot (\autoref{sub:prototype}), and a five-day field study to evaluate their effect on user behavior (\autoref{sub:rq2-method}).

\subsection{Selection of UI Interventions}
\label{sub:intervention-selection}


Intervention design was grounded in choice architecture and persuasive technology or digital nudging~\cite{Thal21,Wein16}, and further informed by sustainability HCI evidence showing that low-energy defaults/options, per-action feedback, pre-action cost visibility, and relatable framing can shift behavior while preserving autonomy~\cite{Gron23,Fili23}.
We selected interventions that act at the moment of interaction, fit the normal chat flow with minimal overhead, provide understandable energy cues, preserve user autonomy, and rely only on runtime-available data. 
We excluded mechanisms, such as social comparison or gamification, because they require additional infrastructure and raise privacy/overhead concerns.


%
Using these criteria and baseline survey receptiveness, we instantiated five interventions. For each, we state \emph{what it is}, \emph{where it comes from}, and the \emph{survey signal} that supported including it:
\begin{itemize}
    \item \threeModeSwitch{}. A control that lets users pick an energy-consuming mode for the chat. It is motivated by the \emph{low-energy configuration} mechanism in digital nudging~\cite{Wein16}, in line with eco-mode and restructured-choice UIs in prior sustainability work~\cite{Gron23,Fili23}. Participants were interested in a lower-energy chatbot mode (Q18--19).
    \item \energyNote{}. A short estimate of energy use shown with each LLM reply. It is motivated by the \emph{contextual feedback} mechanism~\cite{Wein16}, similar to per-action energy or CO\textsubscript{2} feedback in charging interfaces~\cite{Fili23}. 
    Participants wanted usage-level energy information and thought it could change behavior (Q20--21).
    \item \promptPrediction{}. An estimate of energy use shown before the user sends a prompt. The same \emph{contextual feedback} idea~\cite{Wein16}, but placed at the moment of choice, as in pre-decision cost cues~\cite{Fili23}. Participants showed interest in energy information at the point of use (Q20--21).
    \item \metricsDashboard{}. A view that sums energy use over recent activity. It is motivated by the \emph{energy tracking} mechanism~\cite{Wein16}, similar to history-style consumption displays in prior eco-feedback work~\cite{Gron23}. Participants wanted an overview of their overall usage (Q20--21).
    \item \energyAnalogies{}. Energy amounts reframed in everyday units (\eg phone charging). It is motivated by the \emph{understanding mapping} mechanism~\cite{Wein16}, as in relatable equivalencies used elsewhere~\cite{Fili23}. Participants engaged strongly with the iPhone-charging comparison in the survey (Q14).
\end{itemize}
These interventions form the prototype evaluated in \autoref{sub:rq2-method}.
\subsection{RQ$_1$: \rqIshort}
\label{sub:rq1-method}

\begin{center}
    \begin{rqbox}
        \begin{description}
            \item $RQ_1$: \rqI
        \end{description}
    \end{rqbox}
\end{center}

While energy-related awareness among software developers has received increasing attention~\cite{Jagr17,Wyso24}, little is known about how end users perceive the energy impact of their everyday chatbot use.
To establish this baseline, we conducted an online survey probing users' receptiveness to the underlying concepts of the selected interventions---specifically their openness to eco-mode configurations and energy consumption feedback---since the specific UI designs had not yet been prototyped at this stage.

\subsubsection{Survey Design}
Our exploratory survey is grounded in principles of qualitative inquiry~\cite{Patt14}.
It focuses on users' \emph{perceived} awareness, concerns, and attitudes rather than on factual knowledge of energy consumption metrics.

The questionnaire covered four main areas:
(a) user background and chatbot usage patterns,
(b) perceived awareness of energy consumption and the environmental impact of LLM chatbots,
(c) environmental concern and willingness to act, and
(d) attitudes toward sustainability-oriented UI interventions and related trade-offs.

Participants were not expected to have prior knowledge of sustainability concepts or energy measurement units.
To assess users' mental models and uncertainty regarding energy consumption, the survey included estimation-style questions (\eg relative comparisons and everyday analogies).
These questions were designed to elicit perceived understanding rather than accurate estimates, and explicitly allowed participants to indicate uncertainty (\eg ``I'm not sure / I can't estimate'').
This approach captures gaps, misconceptions, and confidence levels in user awareness rather than evaluating technical correctness.

The survey was open between March 18 and April 18, 2025, and distributed via convenience sampling through academic and professional mailing lists, LinkedIn, and targeted emails to individuals who actively use LLM chatbots.
We included both closed-ended and open-ended questions.
The questionnaire was developed iteratively: two authors drafted an initial version based on our research questions, reviewed it internally, and piloted it with two colleagues outside the research team.
Their feedback prompted revisions to wording and question ordering to improve clarity and accessibility for a general audience.

To analyze open-ended responses, we applied thematic analysis~\cite{Braun06} to the substantive responses received.
Two authors independently read all responses and identified themes, then met to compare and discuss their interpretations.
A final set of themes was agreed upon through collaborative refinement; no conflicting interpretations arose during discussion, and no formal inter-rater reliability metrics were computed given the small number of responses and the collaborative nature of the analysis.
Themes were not mutually exclusive, as respondents often addressed multiple aspects in a single response.

The responses were used to establish a baseline of user awareness and attitudes and to confirm user receptiveness to the intervention concepts selected in \autoref{sub:intervention-selection}.
The survey instrument and anonymized response summary are available in the replication package~\cite{reppackage}.

\subsection{Prototype Implementation}
\label{sub:prototype}

\begin{figure}[t]
\centering
\includegraphics[width=0.75\linewidth]{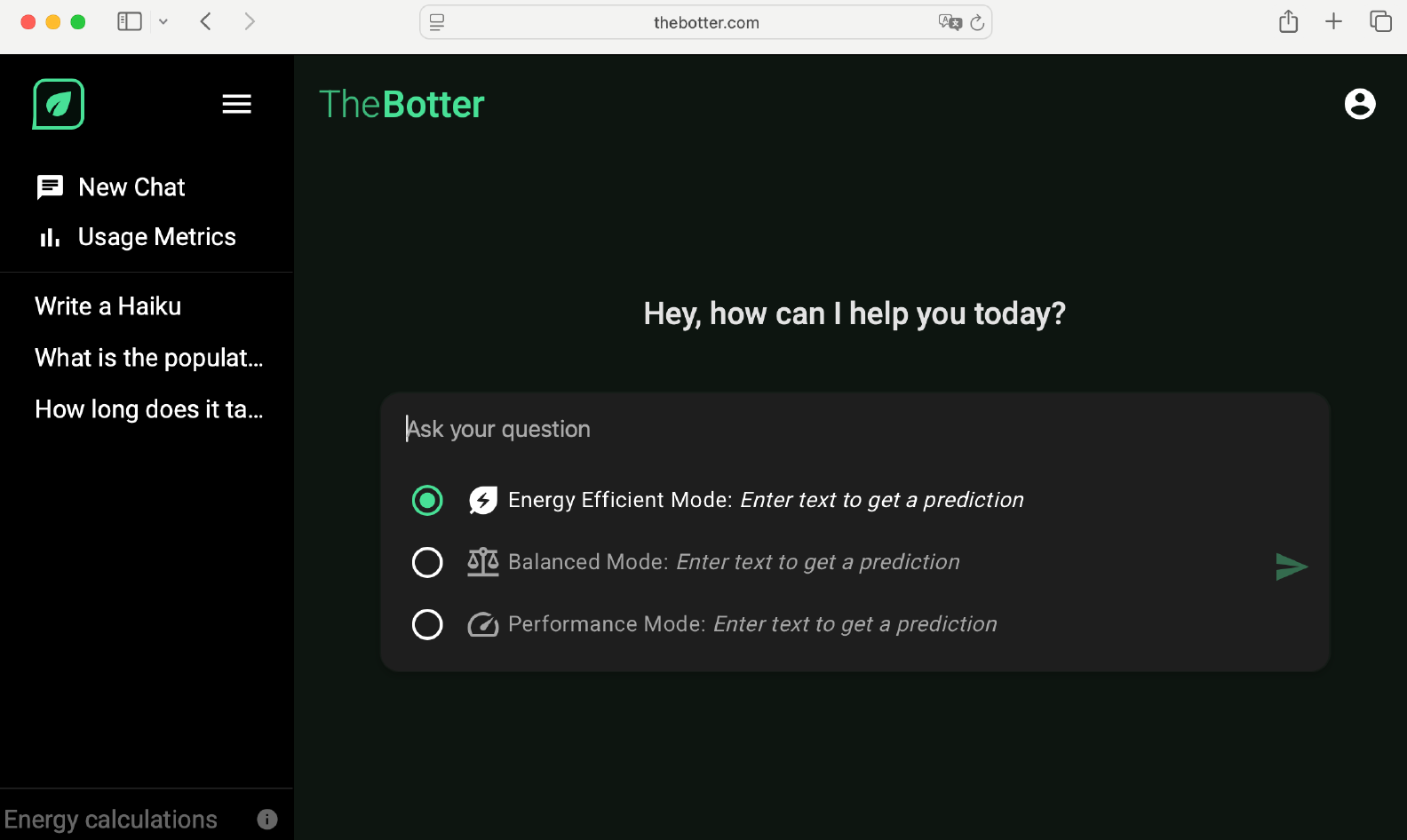}
\caption{\threeModeSwitch{} that lets users select between modes.}
\label{fig:botter-modes}
\end{figure}

\begin{figure}[t]
\centering
\includegraphics[width=0.75\linewidth]{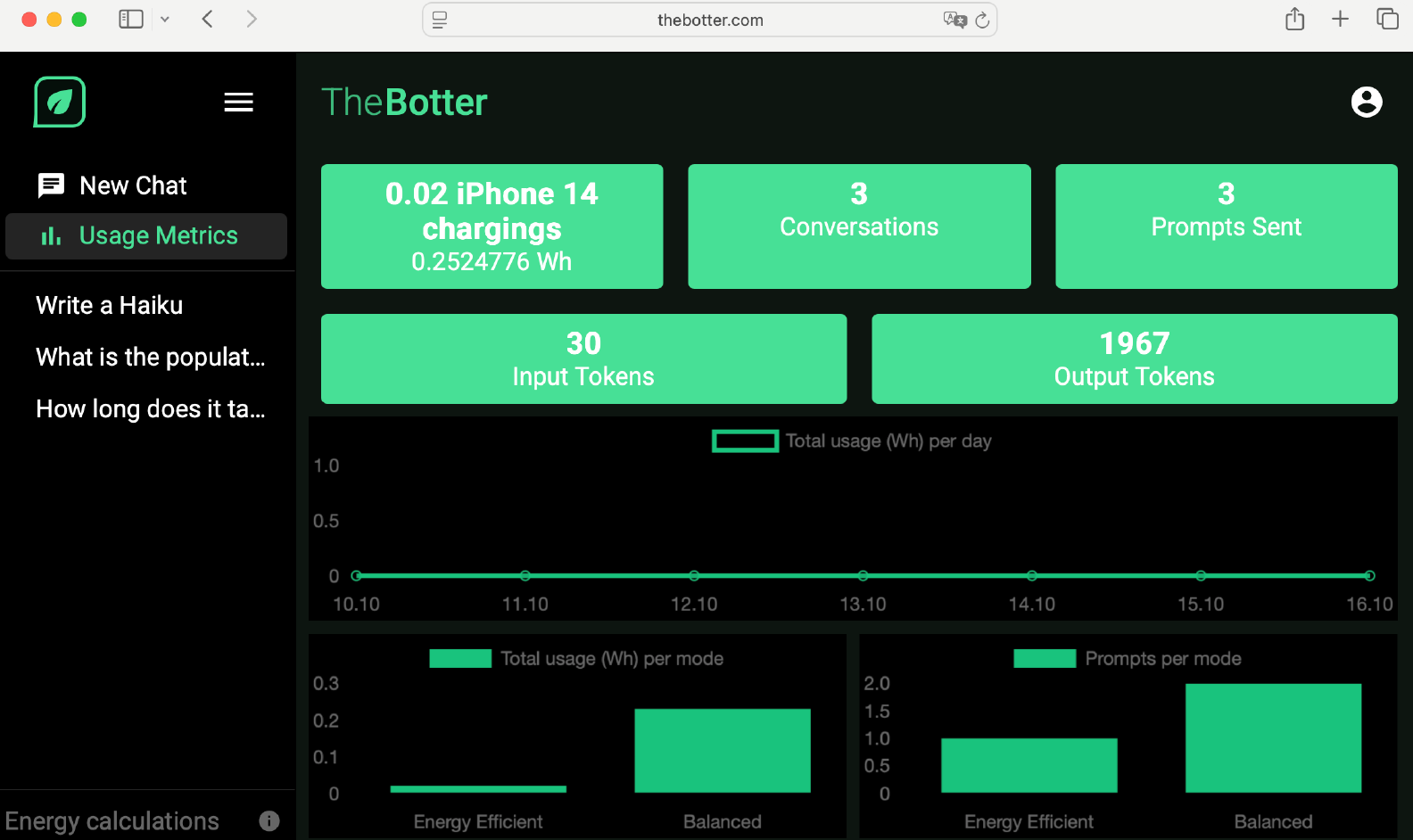}
\caption{\metricsDashboard{} that displays energy consumption.}
\label{fig:botter-dashboard}
\end{figure}

We implemented the five interventions in a chatbot prototype following a client--server architecture.
A Vue.js frontend provides a responsive single-page chat interface, while a Node.js/Express backend manages API communication, intervention state, and energy estimation.
Data---including chat histories, telemetry, and intervention usage logs---is stored in MongoDB.
\autoref{fig:botter-modes} shows the \threeModeSwitch{}, while \autoref{fig:botter-dashboard} presents the \metricsDashboard{}.

We estimated energy use with a simple token-based model, motivated by recent work linking inference energy closely to the number of processed tokens~\cite{Podd25-preprint,Fern25-preprint}. Concretely, the estimated energy per response is:

\begin{equation}
    E = \alpha \cdot T_{\text{in}} + \beta \cdot T_{\text{out}} + \zeta
\end{equation}

\noindent where $T_{\text{in}}$ and $T_{\text{out}}$ are the input and output token counts, and $\alpha$, $\beta$, $\zeta$ are mode-specific coefficients for input cost, output cost, and fixed infrastructure overhead, respectively. 
We assign a higher cost to output tokens in line with benchmark reports and commercial pricing. Coefficients were calibrated against public estimates of GPT-4o query energy~\cite{Jegh25-preprint}, anchoring the model to realistic consumption levels while keeping it lightweight enough for interactive use.
For pre-send prediction (\promptPrediction{}), the number of output tokens is approximated by the input length, which provides stable forecasts in open-ended tasks~\cite{Levy24-preprint}.
To support the \threeModeSwitch{}, coefficients are scaled by relative compute costs published for different model sizes, while the infrastructure overhead remains constant.
The resulting coefficients for each mode are summarized in~\autoref{tab:energy-coefficients}.
\input{energy-coefficients}

Energy estimates are approximations and do not reflect hardware-specific measurements.
Nonetheless, the modular design and transparent parameters make the system reproducible and extensible, supporting future refinement and comparative studies.

\subsection{RQ$_2$: \rqIIshort}
\label{sub:rq2-method}

\begin{center}
    \begin{rqbox}
        \begin{description}
            \item $RQ_2$: \rqII
        \end{description}
    \end{rqbox}
\end{center}

While prior work on AI sustainability has largely focused on model-level improvements such as pruning or hardware efficiency~\cite{Wu22,Tmam24}, the role of UI interventions in shaping energy-related interaction behavior remains underexplored in real-world contexts.
To address this gap, we evaluated the prototype in a five-day field study.

Two research propositions~\cite{Vern09} guided the analysis of RQ$_2$:

\begin{center}
    \begin{rqbox}
        \begin{description}
            \item[P1:] Showing per-prompt consumption and predicted consumption increases the average awareness score from pre- to post-study.
            \item[P2:] Individual end-of-study awareness scores are positively correlated with sustainable behavior logged during the study.
        \end{description}
    \end{rqbox}
\end{center}

\subsubsection{Participants and Setting.}
We recruited participants through the personal and professional networks of the authors.
Eleven individuals voluntarily agreed to participate ($n=11$); no additional filtering criteria were applied beyond willingness to participate and regular use of LLM chatbots in their daily workflows.
Participants used the prototype for five consecutive days on their own devices and within their usual workflows.

\subsubsection{Survey Instrumentation and Telemetry.}
We collected both self-reported indicators and objective usage telemetry:
\begin{itemize}
    \item \textit{Daily awareness check-ins:} Participants completed a short 5-point Likert-scale survey of four questions each day for five days, covering their awareness and the perceived influence of the interventions.
    \item \textit{End-of-the-week questionnaire:} Participants completed 19 questions, including 5-point Likert scale, multiple-select, single-select, and open-ended items, covering themes of awareness and behavioral change. We also evaluated each intervention on understandability, satisfaction, and usability, and collected open-text feedback on what would further raise awareness of energy use.
    \item \textit{System-level telemetry:} For every prompt, the system logged the input and output token counts, selected mode, predicted and actual energy usage, and dashboard visits.
\end{itemize}

All five UI interventions were enabled by default, though participants could disable them at will.

\subsubsection{Study Procedure.}
Participants began with an onboarding awareness questionnaire that introduced them to the chatbot and personalized the interface based on their preferred unit of energy measurement.
Users selected one of five relatable analogies (\eg \textit{iPhone 14 charging}, \textit{minute powering a fridge}).
These units were used consistently across all interventions to enhance comprehension and promote engagement by aligning feedback with familiar reference points.
Over the following five days, participants used the prototype in their regular routines and completed daily awareness check-ins.
At the end of the study, they completed the end-of-the-week questionnaire.
The authors sent a morning reminder to use the prototype and one evening reminder to complete the daily survey.

\subsubsection{Data Preparation and Analysis.}
Survey responses and logs were exported and analyzed together. Methods included:
\begin{itemize}
    \item \textit{Grouped analysis:} Aggregated prompt counts, token usage, and estimated energy consumption per user and per mode.
    \item \textit{Daily trends:} Prompt volume, dashboard visits, and mode distribution across days.
    \item \textit{Prompt growth curves:} Normalized input-edit cycles to study prompt length dynamics.
    \item \textit{Energy efficiency:} Wh per 1{,}000 input tokens per mode, excluding fixed overhead.
    \item \textit{Scatter analysis:} Input vs.\ output tokens with LOWESS regression to assess prediction trends.
\end{itemize}

\subsubsection{Counterfactual energy-savings estimation.}
To quantify how observed behavioral changes translate into potential energy savings, we computed a prompt-level counterfactual using the same logged interactions.
For each prompt $i$, let $E_i^{\text{obs}}$ denote the corrected observed energy estimate (`usageInWhCorrected') under the actually selected mode.
We then defined an all-Performance counterfactual, $E_i^{\text{perf-cf}}$, by re-estimating each prompt as if it had been executed in Performance mode using the same token profile and coefficient model described in~\autoref{sub:prototype}.
The aggregate savings estimate is
\[
\Delta E = \sum_i E_i^{\text{perf-cf}} - \sum_i E_i^{\text{obs}},
\]
and we report both absolute savings (Wh) and relative savings (\% of the counterfactual total).
This estimate should be interpreted as a model-based behavioral counterfactual rather than a hardware-metered causal effect.

\subsubsection{Ethics and Data Handling.}
All participants provided informed consent before participating.
During the study, participants' prompts and session data were temporarily stored on a secure Microsoft Azure server.
Logs, daily check-in responses, and end-of-the-week survey responses are available in the replication package~\cite{reppackage}.

\section{Results}
\label{sec:results}
This section summarizes the findings from the baseline survey (RQ$_1$) and the field user study (RQ$_2$). 

\subsection{RQ$_1$: \rqIshort}
\label{sub:req1-results}

\subsubsection{Survey Participants}
\begin{figure}[!t]
\centering
\subfloat[Agreement levels with AI transparency and energy optimization.\label{fig:stacked:a}]{
    \includegraphics[width=\columnwidth]{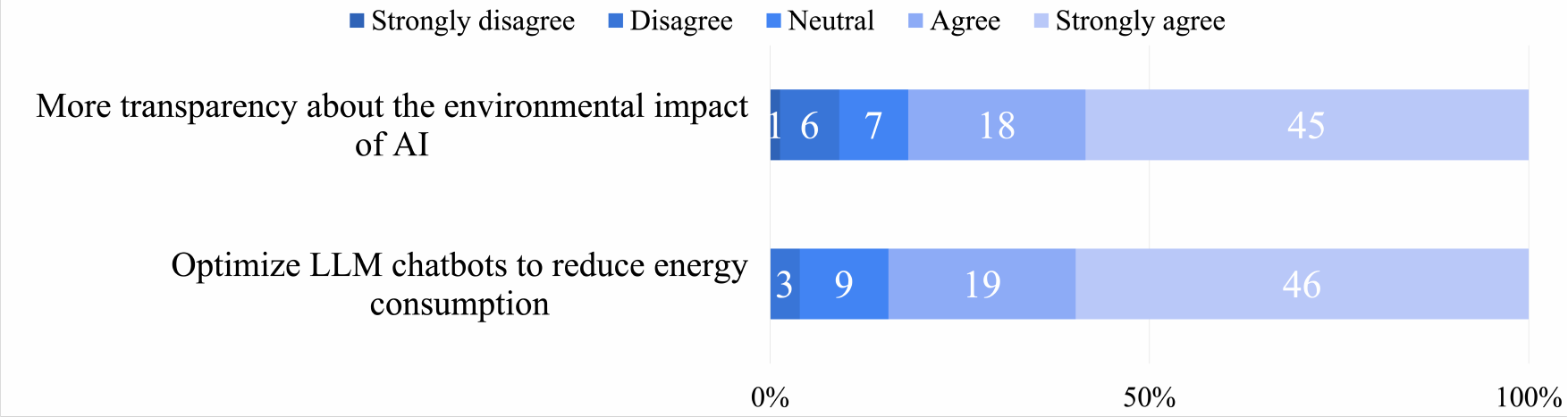}
}\vspace{0.5em}
 
\subfloat[Perceived importance of energy information and environmental impact.\label{fig:stacked:b}]{
    \includegraphics[width=\columnwidth]{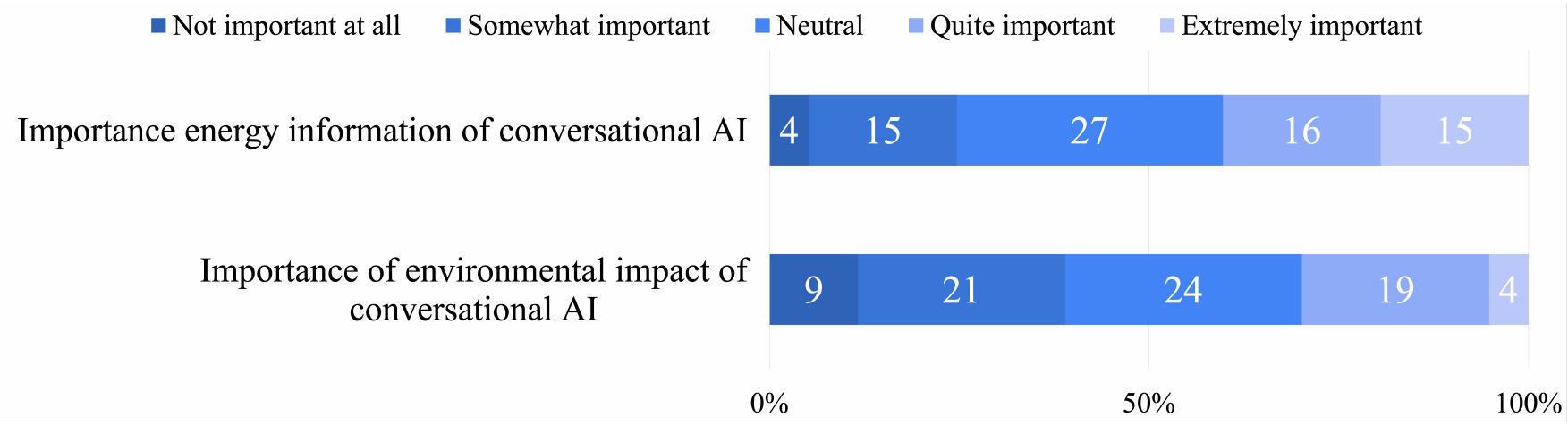}
}\vspace{0.5em}
 
 
\caption{Preferences and Perceived Importance for Sustainable AI Use in Chatbots.}
\label{fig:user-survey-preference-stacked}
\end{figure}

A total of 77 participants completed the survey.
Participants came from diverse professional backgrounds, with academia/research/education (33.8\%, n=26) and developers/engineers (31.2\%, n=24) forming the largest groups, followed by students/interns (16.9\%, n=13).
Most accessed conversational AI tools via desktop or laptop (93.4\%, n=71), with 6.6\% (n=5) using smartphones, and over half (57.9\%, n=44) paid for premium tools. 
Detailed demographic distributions and background characteristics are provided in the replication package~\cite{reppackage}.
~\autoref{fig:user-survey-preference-stacked} summarizes the key results.
 
 
\subsubsection{Usage and awareness}
Most respondents were regular (50.6\%, n=39) or advanced (27.3\%, n=21) users (base: n=76 who answered usage questions); among them, 48.1\% (n=37) used conversational AI tools multiple times per day, and 40.3\% (n=31) several times per week.
 
About 51.9\% (n=40) said they were well aware of LLMs\' energy use; while 42.9\% (n=33) said they were somewhat aware.
Despite claiming high awareness, the estimation tasks showed over-- or underestimation of LLM energy use.
For instance, when comparing a ChatGPT query with a Google search, 48.1\% (n=37) estimated 5$\times$ more energy, 32.5\% (n=25) estimated 10$\times$, and 7.8\% (n=6) estimated 2$\times$; a few said equal (n=2, 2.6\%), less (n=2, 2.6\%), or did not know (n=5, 6.5\%). 
In reality, a ChatGPT prompt consumes roughly 0.30-0.40~Wh, compared to about 0.04~Wh for a Google search--nearly an order of magnitude higher.\footnote{https://blog.samaltman.com/the-gentle-singularity} 
Likewise, for the question ``how many ChatGPT prompts equal one iPhone~14 full charge,'' responses were scattered: 26.0\% (n=20) answered 100 queries, 24.7\% (n=19) answered 50, 22.1\% (n=17) answered 20, 20.8\% (n=16) answered 1{,}000, and 6.5\% (n=5) chose 10{,}000---none of which is the correct answer of approximately 30--40 prompts per full charge. 
\finding{Despite reporting high awareness and frequent use of chatbots, 88.3\% (n=68) of users misestimated their energy cost, revealing their unawareness and \emph{knowledge-perception discrepancy gap}~\cite{Pron07}.}
 
Environmental concern was generally high: only 3.9\% (n=3) expressed no concern about the environmental impact of technology. 
Most showed at least moderate concern, with 18.2\% (n=14) being extremely concerned, 26.0\% (n=20) very concerned, and 24.7\% (n=19) moderately concerned. 
Yet, this concern did not always translate into action.
Although 45.5\% (n=35) indicated that access to energy data would influence how they use AI systems, only 29.9\% (n=23) supported implementing usage limits based on energy impact, compared to 32.5\% (n=25) who opposed and 37.7\% (n=29) who remained unsure. 
Similarly, only 31.2\% (n=24) were in favor of carbon offset payments, with 24.7\% (n=19) undecided and 44.2\% (n=34) opposed.
\finding{Participants expressed strong environmental concern but showed reluctance towards restrictive or compensatory measures, such as usage limits or offset payments, revealing a known \emph{attitude-behavior gap}~\cite{Koll02,Farj19}.}

\subsubsection{Preferences, trade-offs, and decision friction}
Most respondents favored optimizing LLM chatbots for lower energy use, with a strong majority in agreement (84.4\%, n=65).
Most also supported transparency about chatbot's energy consumption from the owner companies: 58.4\% (n=45) strongly agreed, while 23.4\% (n=18) agreed.
40.3\% (n=31) wanted energy-use information, 35.1\% (n=27) were neutral, and 24.7\% (n=19) said no.
 
An ``Energy Efficient Mode'' (\ie \threeModeSwitch{}) was favored by 68.8\% (n=53), with 15.6\% (n=12) unsure and 15.6\% (n=12) not interested.
When explicitly asked about taking a performance trade-off, 39.0\% (n=30) would choose an energy-efficient model even with lower performance, 29.9\% (n=23) were unsure, and 31.2\% (n=24) would not.
 
\finding{Most participants supported optimizing LLM chatbots for energy efficiency, yet willingness declined when personal trade-offs were involved. While 68.8\% favored an Energy Efficient Mode, 31.2\% would not accept lower performance for a smaller footprint. This pattern reflects the \emph{attitude-behavior gap}~\cite{Koll02,Farj19} and \emph{decision friction} under conflicting goals~\cite{Tver92,Dhar97,Luce98}.}
 
\subsubsection{Analysis of open-ended questions}
In response to the open-ended baseline survey question on \textit{how conversational AI could be optimized for sustainability?}, the 77 participants expressed a range of concerns and proposals.
We identified eight recurring themes; their frequencies and representative quotes are summarized in~\autoref{tab:thematic-analysis}.
\input{table-thematic-analysis}

The most frequently mentioned themes were \textit{Infrastructure and Green Energy} (46\%), \textit{Alternative Solutions} (29\%), and \textit{Model and Algorithm Efficiency} (25\%), reflecting a predominant focus on system-level and policy-level changes rather than user-facing design.
Roughly one in four responses touched on \textit{Policy, Taxation, and Offsets} and about one in five on \textit{Pricing and Responsibility}, suggesting that participants see sustainability as a shared responsibility across providers, policymakers, and users.

While these themes signal growing concern about AI's environmental impact, most fall outside the scope of UI design.
The only theme directly relevant to our intervention design was \textit{Awareness and Transparency} (17\%): participants suggested that systems should ``show energy consumption metrics with every query,'' reinforcing the inclusion of per-response energy feedback and a usage dashboard in our prototype.

\subsection{RQ$_2$: \rqIIshort}
\label{sub:req2-results}

The field study involved 11 frequent LLM users over five days. We report descriptive trends (not inferential tests) due to sample size.
Across 278 prompt-response pairs (683{,}580 tokens), estimated observed energy was about 165~Wh.
Under an all-Performance counterfactual, the same interactions would total about 309~Wh.
Observed mode choices were therefore associated with 144.72~Wh lower estimated use (46.77\% reduction). These are model-based counterfactual estimates, not hardware-metered causal effects.

\subsubsection{Behavioral Usage Patterns}
\begin{figure}[!t]
    \centering
    \includegraphics[width=0.85\columnwidth]{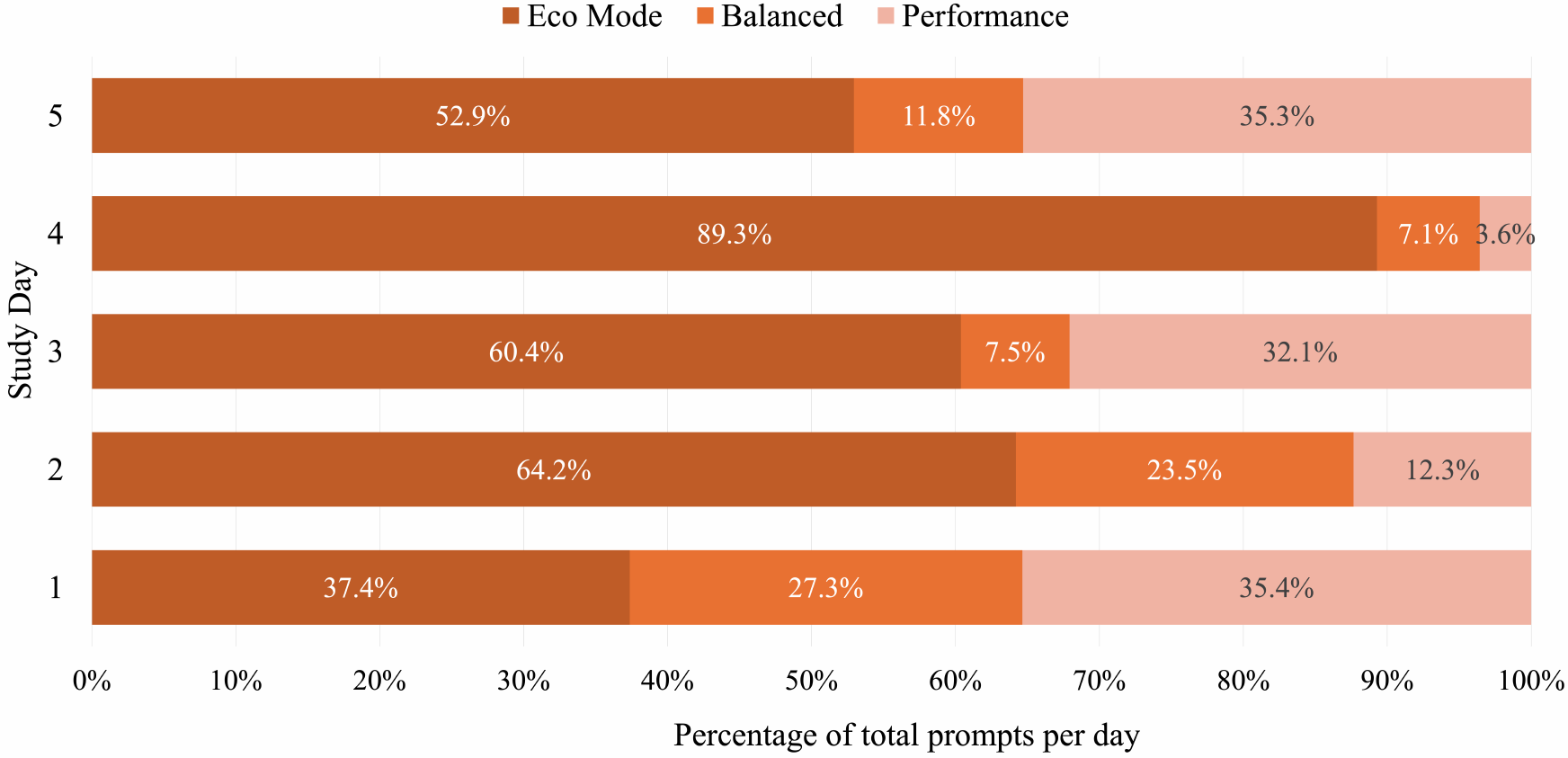}
    \caption{Total prompts sent per mode per day}
    \label{fig:total-prompts-sent-per-mode-per-day}
\end{figure}

\autoref{fig:total-prompts-sent-per-mode-per-day} shows total prompts sent per day per mode.

Participants used the Energy-Efficient mode most frequently (55\%), followed by Performance (25\%) and Balanced (19\%).
Participants viewed the dashboard 41 times in total (an average of 3.7 visits each), with most activity occurring on the first day.
Three heavy users accounted for 77\% of the total footprint.
Prompt lengths remained stable, and scatter plots showed that output tokens frequently exceeded input tokens; LOWESS regression confirmed that the prediction function consistently underestimated outputs.

\subsubsection{Awareness and Behavioral Influence Over Time}
We observed that awareness increased over the course of the study, rising from Day~1 ($M=3.27$, n=11) to Day~5 ($M=4.44$, n=9), with a dip midweek (Days~3--4: $M \approx 3.1$, n=11 and n=10 respectively).

During daily check-ins, participants rated the \energyNote{} and the \threeModeSwitch{} as the most influential interventions (overall $M=3.19$ each across all days).
Dashboard visits occurred on 59.6\% of participant-days, with the highest awareness score on Day~5 coinciding with the day of the strongest dashboard use.

In the end-of-the-week survey, 90.9\% of participants (n=10) reported regularly switching to the Energy-Efficient mode when accuracy was not critical ($M=4.18$).
Prompt shortening was rated neutral ($M=3.00$), and log data confirmed no reduction in input length.

\begin{figure}[!t]
    \centering
    \includegraphics[width=0.75\columnwidth]{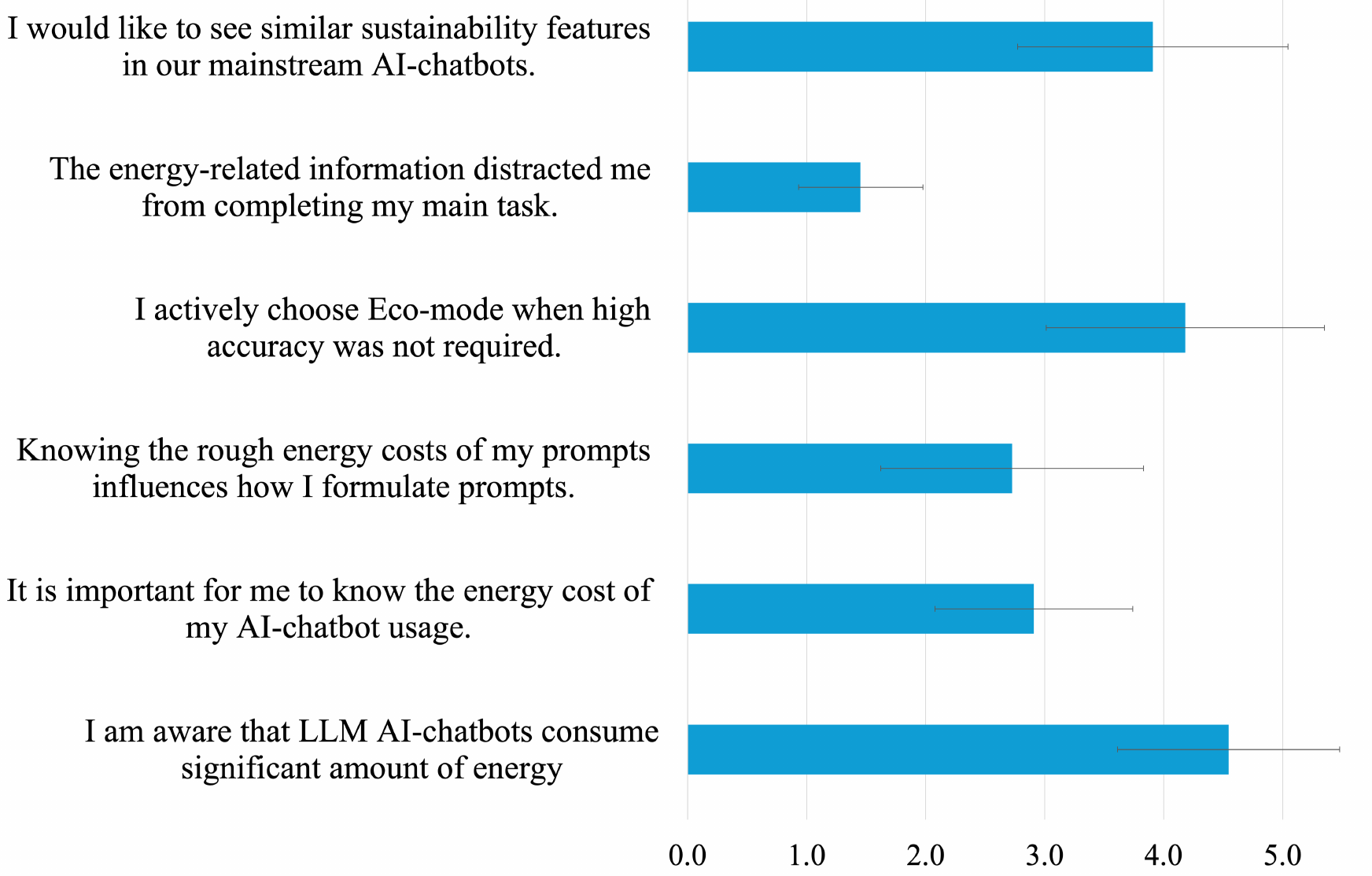}
    \caption{Energy Awareness Trend}
    \label{fig:energy-awareness-trend}
\end{figure}

\autoref{fig:energy-awareness-trend} shows that participants reported high awareness of the energy consumption of LLM chatbots and expressed strong support for sustainability-oriented features.
Most respondents indicated that knowing the energy cost of their prompts influenced how they formulated them and that they actively chose an energy-efficient mode when high accuracy was not required.
At the same time, energy-related information was generally not perceived as distracting.

\finding{Energy awareness rose over time, driven by repeated, low-effort interactions. Participants rated the \energyNote{} and \threeModeSwitch{} as most effective, indicating that \emph{situated, lightweight feedback} can close the knowledge-perception gap and sustain engagement.}

\subsubsection{Usability and Intervention Preferences}
\begin{figure}[!t]
    \centering
    \includegraphics[width=0.75\columnwidth]{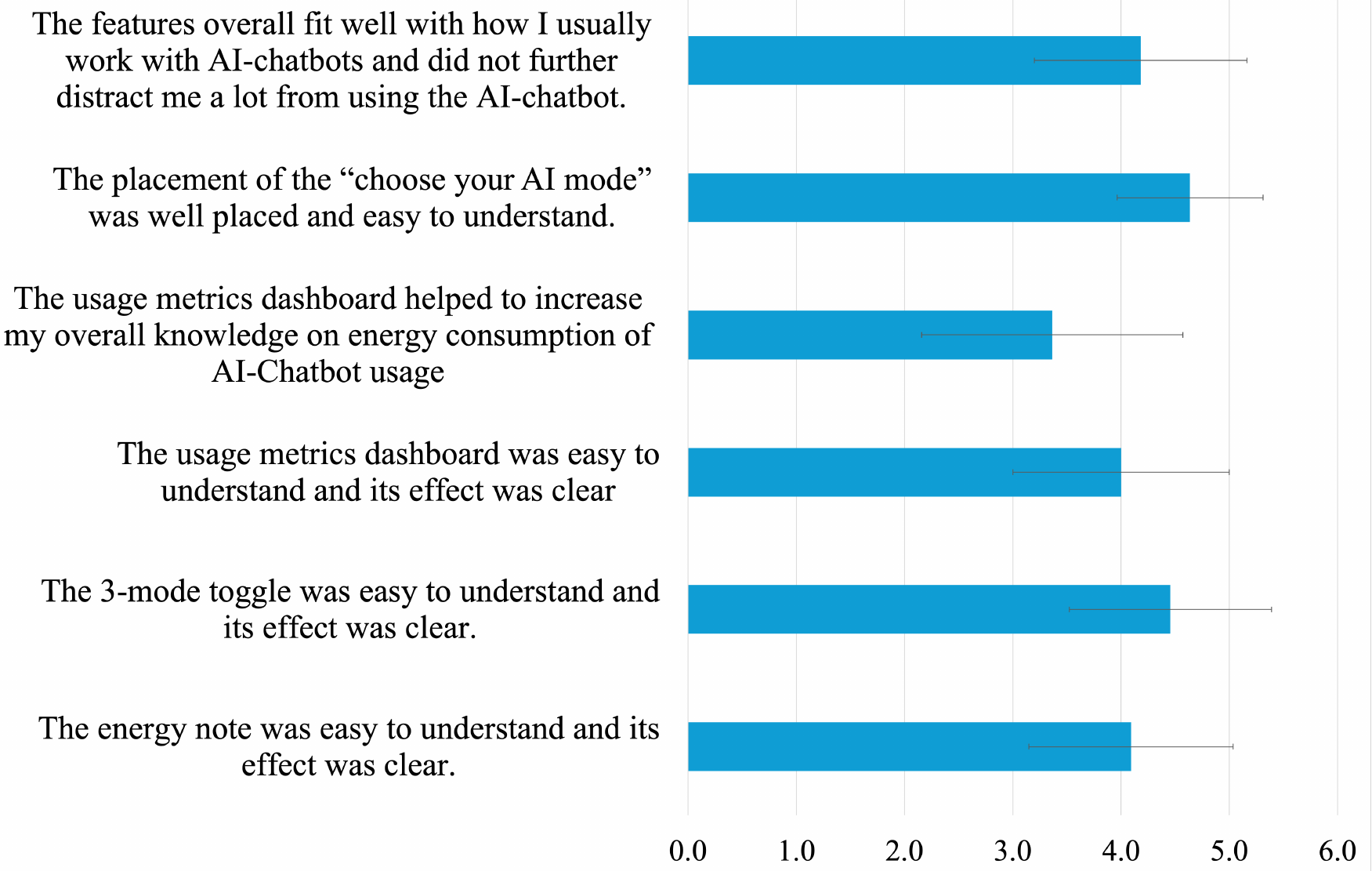}
    \caption{Usability of Interventions}
    \label{fig:interventions_usability}
\end{figure}

All five interventions scored above 4/5 in usability, with the \threeModeSwitch{} rated highest on placement and ease of understanding ($M=4.64$), followed by the \energyNote{} ($M=4.09$) and the \metricsDashboard{} ($M=4.00$), see~\autoref{fig:interventions_usability}.
Qualitative feedback highlighted requests for proactive reminders, aggregate figures in familiar units, and contextual framing over time.

\subsubsection{Energy Impact and Trade-offs}
Although Performance mode was used in less than one quarter of prompts, it was responsible for nearly 89\% of the total energy footprint.
When normalized per 1,000 input tokens, energy consumption was lowest in Energy-Efficient mode (0.064 Wh), moderate in Balanced mode (0.125 Wh), and highest in Performance mode (1.682 Wh).

\begin{center}
    \begin{rqbox}
        Regarding the research propositions~\cite{Vern09},
        P1 (awareness increases over time) was supported by an increase in
        awareness scores from 3.27 to 4.44 over five days, and P2 (awareness
        correlates with sustainable behavior) was supported by the association
        between higher awareness scores, frequent use of Energy-Efficient mode,
        and lower per-user energy footprints.
        Given the small sample size, these observations should be interpreted
        as indicative trends rather than statistically confirmed findings.
    \end{rqbox}
\end{center}

\subsubsection{Qualitative Feedback}
Four participants provided open-ended suggestions. Raw numbers alone felt insufficient: one wrote that ``pure energy consumption figures don't really help me that much---I need something that speaks a bit more to the heart,'' and suggested milestone-style messages (\eg after every 100 prompts) to make cumulative impact tangible. Visibility in the chat mattered: ``I am not aware of my usage on the chat-view \ldots I would like to see current usage in the top bar'' and ``I would understand the usage metrics 100\% if I saw it in terms of my chosen unit.'' Finally, participants asked for lightweight guidance---for example recommending a web search for very short prompts---with any nudges kept ``infrequent and non-intrusive.''

\finding{Users valued transparency but often struggled to interpret raw energy figures meaningfully. This suggests that \emph{contextual and relatable framing} (\eg analogies, milestone messages, and persistent unit display) can reduce decision friction and help users translate awareness into action during everyday use.}

Taken together, the RQ$_2$ results suggest that sustainability-oriented UI interventions can increase awareness and ease routine, low-stakes choices.
However, when users judge accuracy or reliability to be critical, high-performance options remain attractive, and those interactions can still account for a disproportionate share of energy use.
UI support is therefore best understood as a complement to backend and model-level optimization, not a substitute for it.

\section{Discussion, Implications, and Future Work}
\label{sec:implications-future-work}

Our findings align with prior sustainability and choice-architecture research while extending it to LLM chatbots.
The gap between environmental concern and behavior mirrors the attitude-behavior gap~\cite{Koll02,Farj19}, and the strong effect of \threeModeSwitch{} is consistent with evidence that choice-restructuring (eco-mode or reconfigured options) is often more behavior-shifting than information alone~\cite{Gron23,Fili23,Muns16}.
Informational interventions (\energyNote{}, \promptPrediction{}, \metricsDashboard{}) mainly supported awareness over time, while \threeModeSwitch{} drove the clearest mode-selection changes.
At the same time, Performance mode's disproportionate footprint (89\% with <25\% of prompts) reinforces that UI interventions should complement backend and model-level optimization~\cite{Jian24,Wu22}.

This study shows that unsustainable use of LLM chatbots is not primarily driven by a lack of environmental concern, but by everyday decision-making frictions during interaction.
Although most users expressed positive sustainability attitudes, their actual behavior revealed a persistent knowledge-perception and attitude-behavior gap -- a pattern consistent with findings in broader sustainability contexts~\cite{Koll02,Farj19}.

Our results indicate that sustainability-oriented UI interventions are most effective when they reduce effort and decision conflict at the point of choice.
Participants responded more positively to lightweight, inline mechanisms that embedded sustainability reflection directly into the interaction flow, rather than to standalone dashboards or reports.
The \threeModeSwitch{} and per-response \energyNote{} simplified complex trade-offs between accuracy, speed, and energy use into clearly labeled, reversible options.
This reduced hesitation between competing goals and encouraged experimentation without requiring long-term commitment, thereby lowering the perceived cost of acting sustainably.

These findings reinforce the role of decision conflict as a key barrier to sustainable technology use.
When users must balance performance, reliability, and environmental impact, they often default to familiar or high-performance options.
By making trade-offs visible yet easy to revise, the interface supported reflection while preserving user autonomy.
This suggests that sustainable-by-design AI systems should prioritize low-friction choice architectures that support informed decisions without imposing constraints or moral pressure.

At the same time, increased awareness and adoption of energy-efficient modes do not guarantee net reductions in energy consumption.
A key limitation is the potential for rebound effects: users may compensate for weaker or slower responses by switching to more powerful models, issuing additional prompts, or repeating tasks.
In some cases, this behavior could lead to higher overall energy use than selecting a higher-capacity model from the outset.
Moreover, energy-efficient configurations may not be suitable for all tasks, particularly those requiring high accuracy, reliability, or time sensitivity.
These results highlight that UI interventions should be seen as complementary to backend optimizations, model efficiency improvements, and system-level defaults, rather than as standalone solutions.

Future work should therefore study sustainability-oriented UI interventions with larger and more diverse populations and over longer deployment periods.
Such studies are needed to observe persistence, fatigue, habituation, and rebound behaviors over time.
Beyond chatbots, the proposed mechanisms can be transferred to other AI-based systems that expose performance-cost trade-offs, such as recommender systems, search tools, or productivity applications.
Our findings also carry implications for commercial LLM providers. 
While systems such as ChatGPT and Claude already expose model-switching to users, the choice is currently framed around performance and cost, with no visibility into energy consequences. Our results suggest that reframing this choice around energy consumption — and going further, by making mode recommendations adaptive based on task type, prompt complexity, or prior chat history — could meaningfully extend the reach of energy-aware interaction without requiring users to develop expertise in energy estimation. 
For instance, a system that detects a simple factual query and proactively suggests the energy-efficient mode, or that surfaces aggregate footprint data after a session, could lower the friction of sustainable choice-making at scale.
Finally, improving the accuracy and transparency of energy-estimation models remains essential to support informed decision-making and to assess the net environmental impact of these interventions.

\section{Threats to Validity}
\label{sec:threats}

\subsection{Construct Validity}
Construct validity concerns whether our instruments adequately capture the intended concepts.
User awareness and sustainability attitudes were measured through self-reports, which are inherently subject to bias and social desirability effects.
To mitigate this, we piloted the survey to ensure clarity and neutral wording, and we aligned items with constructs commonly used in prior sustainability and awareness research.

We did not expect participants to have prior knowledge of sustainability or energy monitoring, and we did not test such expertise in the survey.
Responses therefore capture perceived awareness and attitudes rather than objective understanding.
This matches our goal of studying how design interventions shape perceptions and behavior, but it limits how strongly we can argue about the factual accuracy of participants' energy-related beliefs.

Some items used estimation-style prompts (\eg relative comparisons or everyday analogies) to elicit mental models and uncertainty.
Participants could indicate uncertainty explicitly, yet items may still have been interpreted in different ways, which can affect response consistency.

Energy use was approximated from token counts.
That choice supports transparent relative comparisons, but absolute values are less reliable given limited telemetry from the model provider.

Several survey and study items used 5-point Likert scales; we report means alongside counts and percentages.
Strictly speaking, Likert items yield ordinal data for which arithmetic means are not well-defined~\cite{Jamieson04}.
Where means appear, they should be read as descriptive summaries of central tendency rather than interval-level measurements.

\subsection{Internal Validity}
Internal validity concerns the extent to which observed effects stem from the interventions themselves.
Our design cannot fully rule out novelty and learning effects: participants may respond positively to a new prototype, or their behavior may shift as they learn the tool.
External influences, such as prior familiarity with sustainability, may also matter.
Even though mode interactions follow a controlled sequence, personal expectations could shape experiences, and background factors (\eg gender, education, experience) could influence awareness and attitudes.
We mitigated these risks by collecting demographic data, qualitatively reviewing participants' comments for confounding patterns, providing an onboarding period that participants could explore without time pressure before logging began, and designing the interface to resemble ChatGPT in layout and visuals while introducing only the sustainability-oriented elements.

Demand characteristics and confirmation bias~\cite{Orne62} are additional concerns.
Because we asked participants to rate energy awareness daily, check-ins may have primed reflection on energy beyond what the UI alone would trigger, potentially inflating self-reported awareness independently of the interventions.
Participants may also have answered in socially desirable ways, a known risk in self-report studies~\cite{Podsakoff03}.
We partially mitigated these concerns by triangulating self-reports with behavioral telemetry (\eg mode selection frequency and dashboard visits), which is less exposed to the same social desirability pressures.

A related concern is hypothesis guessing: because the interventions explicitly frame eco-mode as the low-energy option, participants may have inferred the study's intent and selected it more often than they would in unconstrained use, meaning behavioral telemetry may also be partially shaped by demand effects.
We cannot fully rule this out.
However, Performance mode usage remained non-trivial (25\% of prompts, accounting for $\approx$89\% of footprint), which suggests that participants did not uniformly conform to a perceived ``correct'' eco behavior and continued to make task-driven mode choices.

These precautions strengthen internal validity, but the short study duration and limited diversity among participants still warrant caution when attributing changes solely to the implemented interventions.

\subsection{External Validity}
The generalizability of our findings is limited.
Our sample of 77 participants skews toward technical or academic backgrounds, so attitudes and behaviors may not represent broader populations, and idiosyncratic actions can appear as stable patterns.

The prototype resembled ChatGPT but remained a simplified research system.
Real-world chatbot use involves more varied tasks, longer engagement, and stronger cost or performance trade-offs.
The energy coefficients we applied are also model-specific, so quantitative results may differ across providers and model versions.

Together, these factors limit how far our conclusions extend beyond the studied setting.

\subsection{Conclusion Validity}
The field study's limited sample size reduces statistical power, makes subtle effects hard to detect, and increases the risk of both Type~I and Type~II errors.
For that reason, we emphasize descriptive trends and consistency across participants rather than inferential statistical tests.

Accordingly, changes such as increases in self-reported awareness or shifts in logged interaction behavior should be interpreted as indicative rather than statistically conclusive.
Triangulating survey responses with behavioral metrics strengthens our reading of the evidence, but several conclusions remain tentative given the exploratory nature of the qualitative analysis and the small \(n\).

Interpreting behavioral indicators (\eg prompt-length change) is further complicated by uncertainty in predicting output tokens and in approximating energy consumption.
Themes from open-ended responses were identified through collaborative interpretation rather than independent parallel coding, so fine-grained thematic distinctions should be treated cautiously.

Overall, we intend our conclusions as promising indications that merit replication in larger and more diverse studies.

\section{Related Work}
\label{sec:related-work}

\subsection{Environmental Impact and Energy Efficiency of LLMs}
The rapid deployment of large-scale language models has intensified concerns about their environmental footprint across training, deployment, and use. 
Prior work has quantified the life-cycle energy and carbon costs of LLMs, emphasizing that unchecked scaling of models and workloads risks substantial increases in resource consumption~\cite{Jian24}. 
In response, a growing body of research has focused on improving energy efficiency through model-centric and infrastructure-level optimizations, including renewable-aware training schedules~\cite{Li24a}, workload-based performance modeling~\cite{Wilk24a}, and dynamic cluster reconfiguration for energy-efficient inference~\cite{Stoj25a}. 
Additional studies explore heterogeneous scheduling strategies and benchmarking approaches to balance performance and energy use in large-scale deployments~\cite{Nguy24a,Isaz24}. 
Collectively, this work demonstrates that architectural and operational optimizations are essential for mitigating the environmental impact of LLMs, but it predominantly addresses the backend of AI systems, with limited attention to how user interaction patterns contribute to cumulative energy consumption.

\subsection{User-Facing Interventions and Choice Architecture for Sustainable Behavior}
In parallel, research in persuasive technology, eco-feedback, and behavioral decision making has shown that user-facing design interventions can meaningfully influence behavior toward more sustainable outcomes. 
Prior studies demonstrate that feedback mechanisms, visualizations, nudges, and goal-oriented interfaces can support reflection and encourage resource-conscious choices in domains such as household energy use and digital systems~\cite{Darb06,Froe10,Gee19,Walt25}. 
A closely related body of work on choice architecture highlights how the structuring and framing of options, such as defaults, effort reduction, and contextual feedback, can guide decisions while preserving user autonomy~\cite{Muns16}. 
These principles emphasize reducing cognitive and emotional friction rather than relying solely on awareness or motivation.

While such intervention strategies are well established, their application to LLM-based conversational systems remains limited. 
Existing research on sustainable AI largely overlooks the interaction layer, despite the fact that repeated prompts, model selection, and usage patterns directly affect inference workloads and energy demand. 

\subsection{Model-Switching in Commercial Chatbots}
Several commercial LLM chatbots already expose model-switching to end users. 
OpenAI's ChatGPT allows users to choose between general-purpose models (\eg GPT-4o) and reasoning-intensive models (\eg the o1/o3 series), framing the choice around response quality and latency~\cite{OpenAI-o1-2024}. 
Similarly, Anthropic's Claude offers a tiered model family (Haiku, Sonnet, Opus) at different capability and cost levels, with users guided to select based on task complexity and speed requirements~\cite{Anthropic-Claude3-2024}. 
However, in both cases, the trade-off presented to users is framed around \emph{performance and cost}, with no indication of the energy implications of their choice. 
Our work builds on this existing interaction pattern but reframes mode-switching explicitly around energy consumption, providing users with the transparency needed to make environmentally informed decisions rather than purely cost-driven ones.


\section{Conclusion}
\label{sec:conclusion}
This work examined how UI design can support more sustainable use of LLM chatbots.
From a baseline survey ($n=77$), we identified strong interest in transparency but limited energy understanding, then implemented five interventions in a working prototype.
A five-day field study ($n=11$) showed increased awareness, high usability, and frequent use of energy-efficient mode for non-critical tasks.
Overall, UI-level interventions can improve routine energy-aware behavior, but should be combined with backend and model-level optimization to address high-cost interactions.

\section{Data Availability}
\label{sec:data-availability}
An anonymized replication package containing the survey data, usage logs, and analysis scripts supporting this study is available via a private Figshare link for review purposes~\cite{reppackage}. 


%% file: energy-coefficients.tex
\begin{table}[!t]
\caption{Token-Based Energy Coefficients per Chatbot Mode}
\label{tab:energy-coefficients}
\centering
\footnotesize
\setlength{\tabcolsep}{3pt}
\renewcommand{\arraystretch}{1.05}

\begin{threeparttable}
\begin{tabular}{@{}lccc@{}}
\toprule
\textbf{Mode} & $\alpha$ (Wh/tok.) & $\beta$ (Wh/tok.) & $\zeta$ (Wh) \\
\midrule
Performance\tnote{a}       & 0.00021   & 0.00083   & 0.020 \\
Balanced\tnote{b}          & 0.0000252 & 0.0001008 & 0.020 \\
Energy-Efficient\tnote{c}  & 0.0000042 & 0.0000168 & 0.020 \\
\bottomrule
\end{tabular}

\begin{tablenotes}[flushleft]
\footnotesize
\item[] Notes:
\item[a] GPT\mbox{-}4o ($1.0\times$);
\item[b] GPT\mbox{-}4o\mbox{-}mini ($0.12\times$);
\item[c] GPT\mbox{-}4.1\mbox{-}nano ($0.02\times$).
\end{tablenotes}
\end{threeparttable}
\end{table}

%% file: table-thematic-analysis.tex
\begin{table}[t]
\small
\centering
\caption{Thematic analysis of open-ended survey responses ($n=24$). Themes are not mutually exclusive; percentages reflect the proportion of responses in which each theme appeared.}
\label{tab:thematic-analysis}

\begin{tabularx}{\linewidth}{p{2.8cm} p{4cm} r r X}
\toprule
\textbf{Theme} & \textbf{Description} & \textbf{Count} & \textbf{\%} & \textbf{Representative Quotes from Users} \\
\midrule

Infrastructure/Green Energy & Infrastructure improvements including green energy use in data centers. & 11 & 45.8 &
``Host models in data centers powered by renewable energy.'' \\

Alternative Solutions & Using alternative, more efficient tools instead of LLMs where appropriate. & 7 & 29.2 &
``Automate certain requests to just a Google search instead of an LLM.'' \\

Model/Algorithm Efficiency & Technical improvements to AI models and algorithms for efficiency. & 6 & 25.0 &
``Optimize models for fewer parameters and lower compute demand.'' \\

Policy/Tax/Offsets & Policy-based solutions including taxation, carbon offsets, and regulations. & 6 & 25.0 &
``Introduce carbon taxes on AI usage.'' \\

Pricing/Responsibility & Economic mechanisms and responsibility sharing for sustainable use. & 5 & 20.8 &
``Users should pay a surcharge to cover environmental costs.'' \\

Awareness/Transparency & Increased user awareness and transparency about energy consumption. & 4 & 16.7 &
``Show energy consumption metrics with every query.'' \\

Usage Scope/Limitations & Caution about overuse of LLMs and restricting them to necessary contexts. & 3 & 12.5 &
``There should be a limitation of the use of AI for only necessary tasks.'' \\

Bias/Social Impact & Consideration of broader social impacts and fairness issues. & 2 & 8.3 &
``AI use should be limited considering its societal and fairness implications.'' \\

\bottomrule
\end{tabularx}
\end{table}